\documentclass[aps, showpacs, pra,amssymb, amsmath, twocolumn]{revtex4}
\usepackage{amsmath,latexsym,amssymb}
\usepackage{graphicx}
\usepackage{amsmath}
\usepackage{latexsym}
\usepackage{amssymb}
\usepackage{bm}
\usepackage{placeins}
\usepackage{color}

\newcommand{\ket}[1]{\left|#1\right\rangle}

\begin{document}

\title{Selective excitation in a three-state system using a hybrid adiabatic-nonadiabatic interaction}
\author{Yunheung Song, Han-gyeol Lee, Hanlae Jo, and Jaewook Ahn}
\address{Department of Physics, KAIST, Daejeon 305-701, Korea}
\date{\today}

\begin{abstract} 
The chirped-pulse interaction in the adiabatic coupling regime induces cyclic permutations of the energy states of a three-level system in the $V$-type configuration, which process is known as the three-level chirped rapid adiabatic passage. Here we show that a spectral hole in a chirped pulse can turn on and off one of the two adiabatic crossing points of this process, reducing the system to an effective two-level system. The given hybrid adiabatic-nonadiabatic transition results in selective excitation of the three-level system, controlled by the laser intensity and spectral position of the hole as well as the sign of the chirp parameter. Experiments are performed with shaped femtosecond laser pulses and the three lowest energy-levels (5S$_{1/2}$, 5P$_{1/2}$, and 5P$_{3/2}$) of atomic rubidium ($^{85}$Rb), of which the result shows good agreement with the theoretically analyzed dynamics. The result indicates that our method, being combined with the ordinary chirped-RAP, implements an adiabatic transitions between the two excited states. Furthermore the laser intensity-dependent control may have applications including selective excitations of atoms or ions arranged in space when being used in conjunction with laser beam profile programming.
\end{abstract}
\pacs{32.80.Qk, 78.47.jh, 42.65.Re}

\maketitle

\section{Introduction}

Adiabatic control of a quantum system through its adiabatic evolution path allows robust manipulation and high-fidelity state-preparation~\cite{ShoreBook}. Gradually being recognized as an important requirement in quantum information processing~\cite{NielsonBook}, it has been under active investigation in recent years~\cite{Goswami, Zhang, Ichimura, Renzoni, Vitanov, Moelmer, Zheng}. The best known examples of the adiabatic control methods are the rapid adiabatic passage (RAP)~\cite{VitanovRev,Warren1992, Sauerbrey2002} and the stimulated Raman adiabatic passage (STIRAP)~\cite{Bergmann}. In RAP, an excitation pulse with a monotonic frequency sweep induces a state vector to follow a population-inversion path in Hilbert space traced by an adiabatic state. In STIRAP, a pair of excitation pulses separated in time creates a population-trapping state of a three-level system and the state evolution through the subsequent adiabatic path ensures 100\% population transfer from the initial state to the final state without populating the intermediate state. Both of these methods have been widely applied to various quantum systems including  atom optics~\cite{Muga}, NMR~\cite{DelaBarre}, cavity quantum electrodynamics~\cite{Kimble}, superconducting qubits~\cite{Paraoanu}, and quantum dots~\cite{Chatel}.

Broadband light sources greatly benefit optical approaches to qubit manipulations because of their powerful pulse-shape programming capability~\cite{WeinerBook, WeinerOC2011}. In ultrafast optics, composing the amplitude and phase of a broadband laser pulse provides various complex pulse shapes, and their usage often plays a crucial role in investigating and engineering new quantum dynamics of atoms and molecules~\cite{Silberberg, Bayer,  Sangkyung2012, Hangyeol2013, Hangyeol2016, Baumert}. Of particular relevance in the context of the present paper is the selective population method of dressed states (SPODS)~\cite{Baumert} which provides a pulse shaping scheme especially in the frequency domain for strong-field controls of multilevel systems. 

In this paper we consider a laser pulse shaping method to embed a local nonadiabatic coupling in the middle of a three-level chirped RAP process~\cite{Warren1992, Sauerbrey2002}. The chirped RAP makes a set of cyclic permutation transitions for a three-level system in the $V$-type configuration: $\ket{0}\rightarrow \ket{1}$, $\ket{1}\rightarrow \ket{0} \rightarrow \ket{2}$, and $\ket{2}\rightarrow \ket{0}$ (for a positive chirp, and a negative chirp reverses the directions), when $\ket{1}$ and $\ket{2}$ are the excited states and $\ket{0}$ is the ground state $\ket{0}$, and at the first and second adiabatic crossing points the state $\ket{0}$ is interchanged with $\ket{1}$ and $\ket{2}$, respectively. So if the transition at the first adiabatic crossing is turned off (with the new nonadiabatic coupling), the states $\ket{0}$ and $\ket{1}$ bypass the crossing and the states $\ket{0}$ and $\ket{2}$ are interchanged at the second crossing and the state $\ket{1}$ is unchanged (i.e., $\ket{0}\rightarrow \ket{2}$, $\ket{1}\rightarrow \ket{1}$ and $\ket{2}\rightarrow \ket{0}$). We will show that this type of hybrid adiabatic-nonadiabatic interaction can be implemented with a chirped optical pulse with a spectral hole resonant to one of the two excited states. In our method the laser intensity is used to turn on or off the nonadiabatic coupling, while in a conventional RAP approach the spectral chirp sign is used for the selectivity~\cite{Warren1992, Sauerbrey2002}. Using the laser intensity as a control parameter brings along the benefit of spatial beam shaping, which enables simultaneous control of multiple qubits arranged in space (to be discussed as an application). 

The remaining sections are organized as follows: We first theoretically study the model Hamiltonian for the given shaped-pulse interaction with a three-level system in Sec.~II, where we show that the chirped pulse with a spectral hole can embed the non-adiabatic coupling amid a conventional RAP process. After the experimental procedure is briefly illustrated in Sec. III, the experimental results are provided in Sec. IV, where the intensity-dependent selectivity of the as-designed hybrid adiabatic-nonadiabatic interaction is presented. The conclusion follows in Sec. V.

\section{Theoretical Consideration}
The model system is a three-level atom in the $V$-type configuration, consisting of two excited energy states, $\ket{1}$ and $\ket{2}$, and the ground state, $|0\rangle$ (of respective energies $\hbar\omega_1$, $\hbar\omega_2$, and $0$). We consider this system is interacted with a spectrally-shaped laser pulse, a chirped Gaussian pulse with a spectral hole, defined in the spectral domain as
\begin{eqnarray}
E(\omega) &= &  E_0 \Big( e^{-{(\omega-\omega_m)^2}/{\Delta \omega_m^2}}-
\alpha e^{-{(\omega-\omega_h)^2}/{\Delta \omega_h^2}}
\Big) \nonumber \\&& \times 
e^{-i{c_2}(\omega-\omega_m)^2/2},
\label{EComega}
\end{eqnarray}
where $\omega_m$ and $\omega_h$ are the center frequencies of the main pulse and the hole, respectively, $\Delta\omega_m$ and $\Delta\omega_h$ are the the bandwidths, and $c_2$ is the chirp parameter~\cite{WeinerBook}. The condition $\alpha=\exp[{-{(\omega_h-\omega_m)^2}/{\Delta \omega_m^2}}]$ in Eq.~\eqref{EComega} makes a complete spectral hole around $\omega=\omega_h$.
The electric field in the time domain is the inverse Fourier transform of $E(\omega)+E(-\omega)$, which is given by
\begin{eqnarray}
E(t) &=&  \frac{\mathcal{E}_m(t)}{2} e^{i\{(\omega_m +\beta_m t )t + \varphi_m\}} \nonumber \\
&&-\frac{\mathcal{E}_h(t)}{2} e^{i\{\omega_h + \beta_h (t-\gamma)\}(t-\gamma)+ \varphi_h\}}  +  c.c. \nonumber\\
&\equiv&E_m(t)+E_h(t)+c.c.,
\label{Efield}
\end{eqnarray}
where $\beta_m = {c_2}/(2c_2^2+8/\Delta\omega_m^4)$ and $\beta_h = {c_2}/(2c_2^2+8/\Delta\omega_h^4)$ are the chirp parameters for the main and hole pulses, respectively, and $\gamma=-c_2(\omega_m-\omega_h)$ is the time shift of the hole with respect to the main pulse. The amplitudes and (time-independent) phases of the pulses are respectively given by
\begin{eqnarray}
\mathcal{E}_m(t) &=& E_0\sqrt{\frac{\Delta\omega_m}{\tau_m}} e^{-{t^2}/{\tau_m^2}}, \\
\varphi_m &=&-\frac{1}{2}\tan^{-1}\frac{c_2\Delta\omega_m^2}{2},\\
\mathcal{E}_h(t) &=& \alpha E_0\sqrt{\frac{\Delta\omega_h}{\tau_h}} e^{-{(t-\gamma)^2}/{\tau_h^2}}, \\
\varphi_h &=&-\frac{1}{2}\tan^{-1}\frac{c_2\Delta\omega_h^2}{2}-\frac{c_2}{2}(\omega_m^2-\omega_h^2),
\end{eqnarray}
where $\tau_i=\sqrt{4/\Delta\omega_i^2+c_2^2\Delta\omega_i^2}$ is the Gaussian pulse width of each chirped pulse $i=m,h$. 

\begin{figure}[thb]
   \centerline{\includegraphics[width=0.45\textwidth]{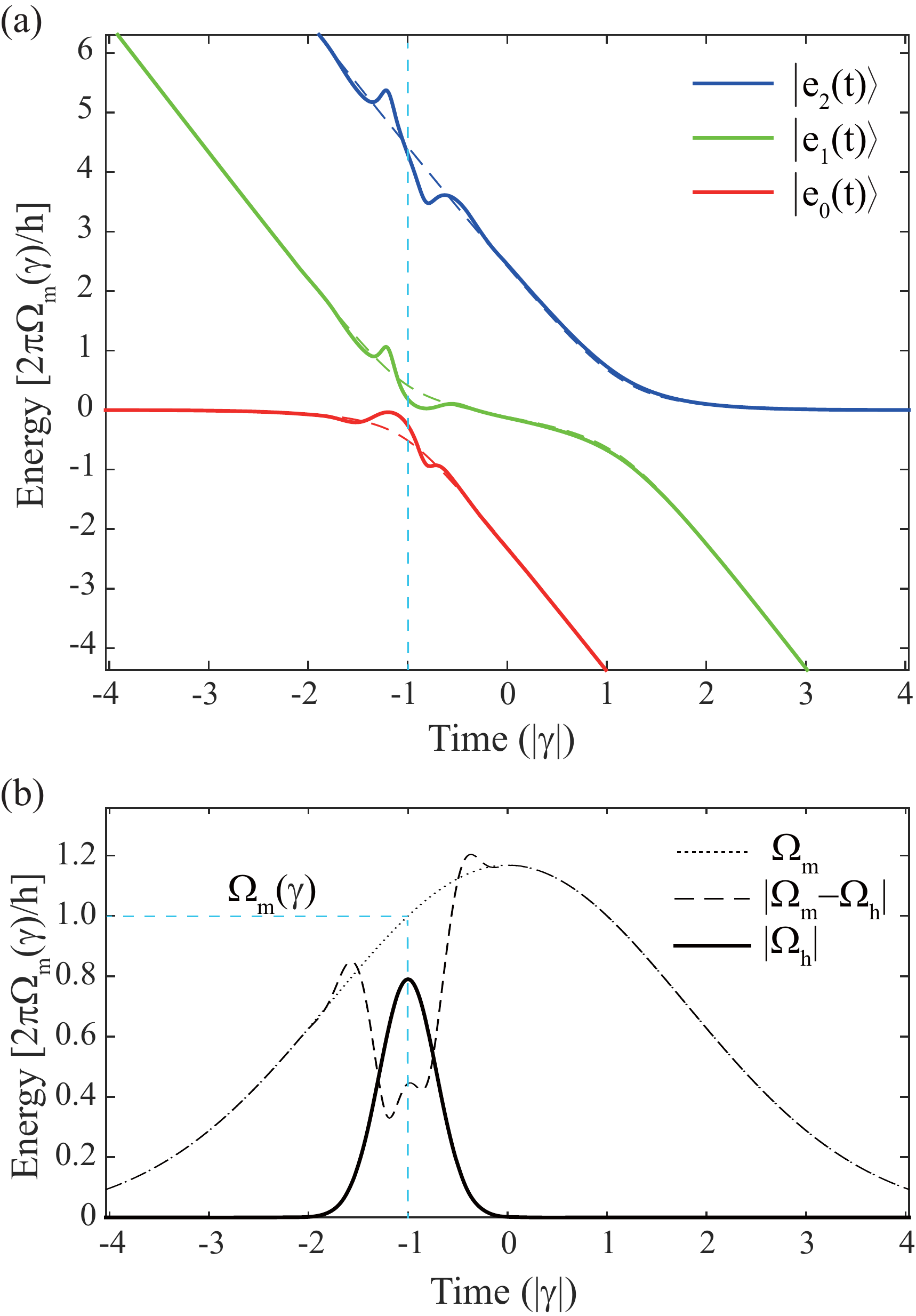}}
    \caption{(Color online) (a) Eigenstate energies of the Hamiltonian including (solid lines) and excluding the hole pulse interaction (dashed lines). (b) The temporal envelopes of the hole pulse (solid line), the main pulse (dotted pulse), and the total pulse (dotted line). }
    \label{fig1}
\end{figure}

Suppose that the main pulse is frequency-centered between the excited states (i.e., $\omega_m=(\omega_1+\omega_2)/2$), with a bandwidth enough to cover the both states (i.e., $\Delta\omega_m > \delta\equiv \omega_2-\omega_1$) and that the hole pulse is resonant to only one of them, say, $\ket{1}$ (i.e., $\omega_h=\omega_1$ and $\Delta\omega_h <\delta$). The Hamiltonian is then given in the eigenstate basis by 
\begin{eqnarray}
H(t) &=& 
\left[ \begin{array}{ccc}
e_0(t) & 0 & 0 \\
0 & e_1(t) & 0 \\
0 & 0 & e_2(t) 
\end{array} \right]
- i\hbar R\dot{R}^{-1}, 
\label{adiaHamiltonian}
\end{eqnarray}
where $\{e_j(t) \}$ are the eigenstate energies and $R_{jk}(t)=\left<e_j|k\right> $ is the transformation from the bare basis $\{\ket{k}\}$ to the eigenstate basis $\{\ket{e_j}\}$ for $j,k \in \{0,1,2\}$. The time evolution of the eigen-energies $e_j(t)$  is plotted in Fig.~\ref{fig1}(a), where we use $\Delta\omega_m=10\Delta\omega_h=2.5\delta$, and a positive $c_2=2/\Delta\omega_h^2$ is chosen to satisfy the minimum hole pulse-width condition for a constant chirp parameter. The main pulse induces slowly-varying adiabatic passages (the dotted lines) and the hole the rapid change of them (the solid lines) near the first adiabatic crossing point. These behaviors are consistent with the temporal profiles of the main pulse and the hole, as shown in Fig.~\ref{fig1}(b). Note that the the instantaneous frequency of the main pulse becomes equal to $\omega_1$ at $t=\gamma$.

{\it (i) Fully adiabatic coupling regime:} When the nonadiabatic coupling term, $-i\hbar R\dot{R}^{-1}$ in Eq.~\eqref{adiaHamiltonian}, is small, the adiabatic condition~\cite{VitanovRev} is satisfied in all time. Each eigenstate $\ket{e_i}$ is an adiabatic state, evolving from one bare state $\ket{i}$ to the next one $\ket{i+1}$ (cyclically), i.e.,
\begin{eqnarray}
\lim_{t \rightarrow -\infty} |e_0(t)\rangle = |0\rangle, &\quad&
\lim_{t \rightarrow \infty} |e_0(t)\rangle = |1\rangle, \nonumber \\
\lim_{t \rightarrow -\infty} |e_1(t)\rangle = |1\rangle, &\quad& 
\lim_{t \rightarrow \infty} |e_1(t)\rangle = |2\rangle,\nonumber\\
\lim_{t \rightarrow -\infty} |e_2(t)\rangle = |2\rangle, &\quad& 
\lim_{t \rightarrow \infty} |e_2(t)\rangle = |0\rangle,
\end{eqnarray}
up to a global phase. (A negative chirp reverses the direction of the three-state cyclic permutations.) So, the result in the fully adiabatic three-state coupling regime is a cyclic transition ($\ket{0}\rightarrow \ket{1}$, $\ket{1} \rightarrow \ket{2}$, $\ket{2} \rightarrow \ket{0}$), similar to the three-level chirped RAP~\cite{Warren1992,Sauerbrey2002}. 

\begin{figure}[thb]
  \centering
  \includegraphics[width=0.45 \textwidth]{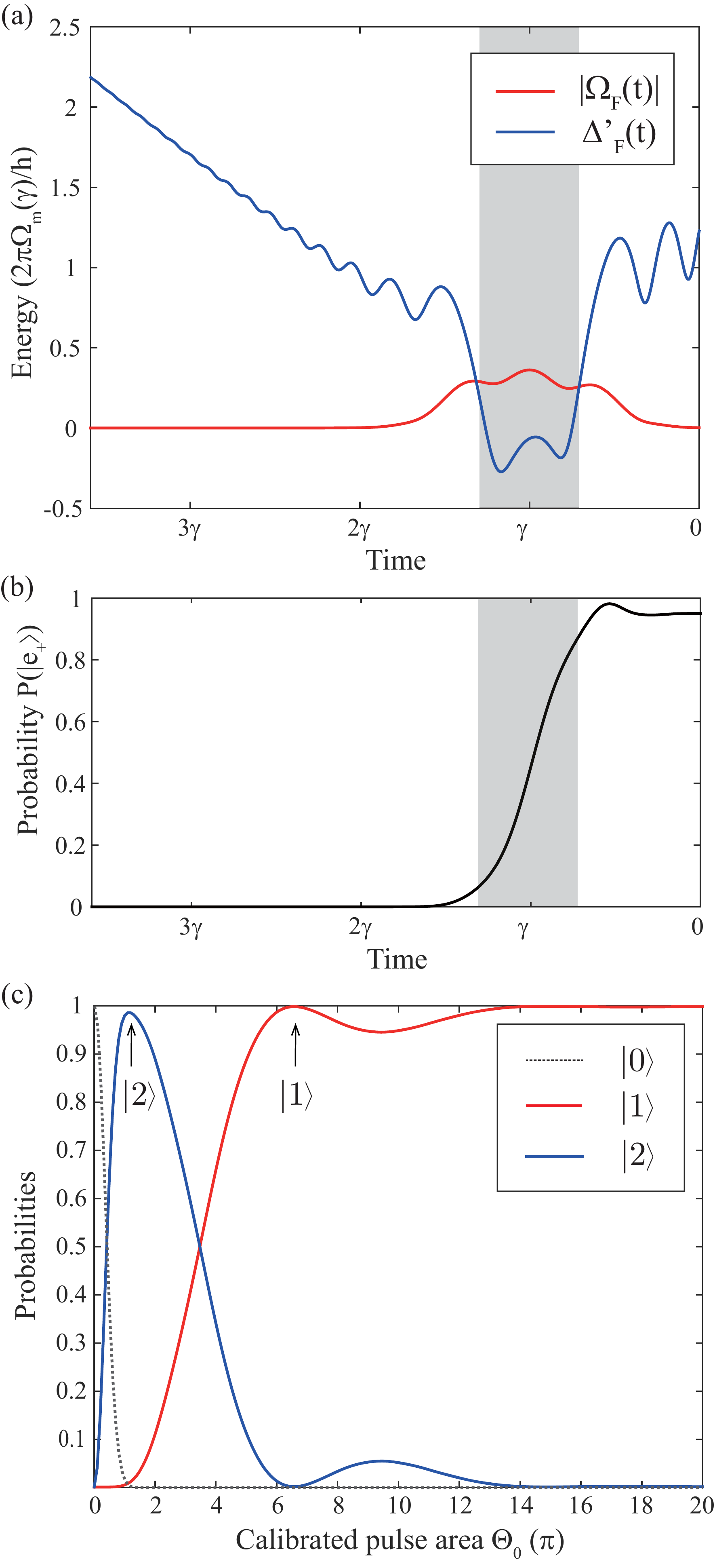}
\caption{(Color online) (a) The effective coupling and detuning, defined in Eqs.~\eqref{eq13} and \eqref{eq14}, of the coupled two-level system. (b) The probability of the adiabatic energy state $\ket{e_+}$. (c) Probabilities of the bare energy states vs. calibrated pulse area $\Theta_0$.}
\label{fig2}
\end{figure}

{\it (ii) Hybrid adiabatic-nonadiabatic coupling regime:} When the adiabatic condition is violated due to the presence of the hole (we consider the main pulse alone is still adiabatic), the given Hamiltonian results in a hybrid adiabatic-nonadiabatic transition between $\ket{e_0}$ and $\ket{e_1}$. Since the nonadiabatic coupling is localized in time near $t=\gamma$, we may consider two separate time regions: $t>0$ and $t<0$. In the positive time region, the dynamics is a fully adiabatic process, so the state at $t=0$ simply remains till $t=\infty$. In the negative time region, the eigenstate $\ket{e_2}$ can be decoupled because it is far-off resonant from $\ket{e_0}$. When we rewrite the Hamiltonian in the adiabatic basis of the main pulse (only), the Hamiltonian is given under the rotating-wave approximation by
\begin{eqnarray}
H'(t<0) &=& 
\left[ \begin{array}{ccc}
e_-(t) & 0 & 0 \\
0 & e_+(t) & 0 \\
0 & 0 & 2\delta+\Delta(t)
\end{array} \right] \nonumber\\
& -& \frac{\hbar}{2} R'\left[ \begin{array}{ccc}
0 & \Omega_{h}(t) & 0\\
\Omega_{h}^*(t) & 0 & 0 \\
0 & 0 & 0
\end{array} \right]R'^{-1},
\label{2leveladiaHamiltonian}
\end{eqnarray}
where $e_{\pm}(t)={\hbar}[\Delta(t)\pm\sqrt{\Omega_{m}(t)^2+\Delta(t)^2}]/{2}$ are the adiabatic energies, $\Delta(t)=-\delta-2\alpha t$ is the detuning (for the main pulse), and $R'(t)$ is the transform matrix given by
\begin{equation}
R'(t)=
\left[\begin{array}{ccc}
\cos{\vartheta(t)} & -\sin{\vartheta(t)} & 0\\
\sin{\vartheta(t)} & \cos{\vartheta(t)} & 0\\
0 & 0 & 1
\end{array}\right],
\end{equation}
with the mixing angle 
\begin{equation}
\vartheta(t)=\frac{1}{2}\tan^{-1}{\frac{\Omega_{m}(t)}{\Delta(t)}}\quad{\rm for}\quad 0\le\vartheta(t)\le\frac{\pi}{2}.
\end{equation}
The Rabi frequency for the transition from $\ket{0}$ to $\ket{1}$ is defined by $\Omega_{i}(t)=2\mu_{01}{E}_i(t)\exp[-i(\alpha t^2+\omega_m t+ \phi_m)]/\hbar$  for each pulse $i=m,h$, where the phase factor of the main pulse is added to keep $\Omega_{m}$ real. 

The Hamiltonian in Eq.~\eqref{2leveladiaHamiltonian} can be simplified to be
\begin{equation}
H_F(t)
= \hbar \left[
\begin{array}{ccc}
-\Delta_F(t)/2 & \Omega_F(t)/2 & 0 \\
\Omega_F^*(t)/2 & \Delta_F(t)/2 & 0\\
0 & 0 & 2\delta+\Delta(t)/2
\end{array}\right] 
\end{equation}
with the effective coupling $\Omega_F$ and detuning $\Delta_F$ of the coupled two-level system ($\ket{e_+}$ and $\ket{e_-}$), defined by
\begin{eqnarray}
\Omega_F(t)&=&-{\rm Re}(\Omega_{h}) \cos 2\vartheta-i {\rm Im}(\Omega_{h}), \label{eq13} \\
\Delta_F(t)&=&\sqrt{\Omega_{m}^2+\Delta^2}-{\rm Re}(\Omega_{h})\sin 2\vartheta. \label{eq14}
\end{eqnarray}
Note that similar coupling and detuning terms are discussed in the context of the zero-area pulse interaction with a two-level system~\cite{Hangyeol2016}.

The phase of $\Omega_F$(t) in Eq.~\eqref{eq13} is time-dependent, so the dynamics can be better explained in the interaction-picture. Figure~\ref{fig2}(a) shows the numerical calculation of the coupling $|\Omega_F(t)|$ and the detuning $\Delta_F'(t)=\Delta_F(t)+d \arg[\Omega_F(t)]/dt$ in the interaction picture. Their plateau region around $t=\gamma$, the first (non-adiabatic) crossing point, manifests a near-resonant two-state coupling, which results in the complete population inversion ($\ket{e_0} \rightarrow \ket{e_1}$, $\ket{e_1} \rightarrow \ket{e_0}$) in the adiabatic basis, as shown in Fig.~\ref{fig2}(b). When the system evolves further to the second crossing point (at which the adiabaticity is satisfied), the state $\ket{e_1}$ continues to remain in $\ket{e_1}$. So the given three-level system results in a closed two-level system, $\ket{0}$ and $\ket{2}$, in the bare-atom basis, plus an isolated state $\ket{1}$.

Figure~\ref{fig2}(c) shows the fully-numerical calculation of the final state populations in the bare-atomic basis using the Hamiltonian in Eq.~\eqref{adiaHamiltonian} which includes the nonadiabatic coupling term, where the calibrated pulse-area $\Theta_0$ is defined with the transform-limited pulse having the same energy of the total pulse. As the pulse energy increases, the state $\ket{0(t=-\infty)}$ either remains in $\ket{0(t=\infty)}$ (in the fully non-adiabatic regime for small $\Theta_0$), or evolves to $\ket{2(t=\infty)}$ (in the hybrid adiabatic-nonadiabatic regime for in-between $\Theta_0$), or to $\ket{1(t=\infty)}$ (in the fully adiabatic regime for large $\Theta_0$). The result indicates the selective transitions to any energy state of the three-level system (i.e., $\ket{0} \rightarrow \ket{0}$, $\ket{1}$, or $\ket{2}$), controlled with only laser intensity (in the hybrid adiabatic-nonadiabatic coupling regime).  

\begin{figure}[tbh]
  \centering
  \includegraphics[width=0.45 \textwidth]{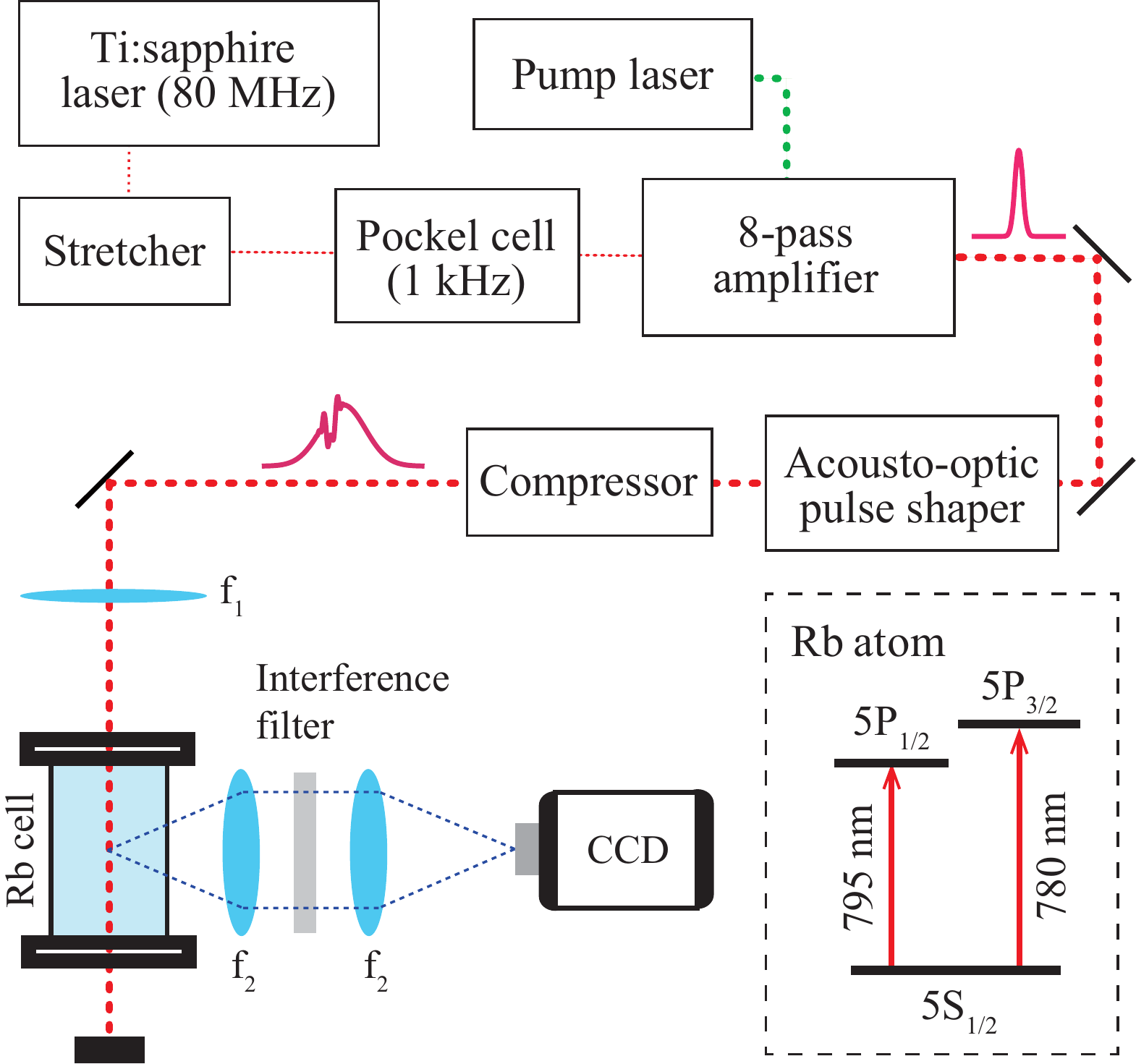}
\caption{(Color online) Schematic experimental setup. }
\label{setup}
\end{figure}

\section{Experimental Procedure}

The experimental setup is schematically shown in Fig.~\ref{setup}.  We used ultrashort optical pulses from a Ti:sapphire laser amplifier (homemade) operating with a repetition rate of 1~kHz and a pulse energy up to 25~$\mu$J. The spectral bandwidth, the full width at the half maximum (FWHM), was 30~nm ($\Delta\omega_m=2\pi\times9~{\rm THz}$). The center wavelength was tuned to $\lambda=787.6$~nm ($\omega_m=2\pi c/\lambda$), the center between the transitions to $5P_{1/2}$ and $5P_{3/2}$ energy levels from the ground state $5S_{1/2}$ of atomic rubidium ($^{85}$Rb). Each laser pulse was programmed with an acousto-optic pulse shaper~(AOPDF, Dazzler)~\cite{AOPDF}. The spectral hole was centered at the transition to $5P_{1/2}$, and the linear chirp was varied from $c_2 = -20,000$ to $50,000$~fs$^2$ by the AOPDF.  The shaped laser pulses were focused with an $f_1=1000$~mm lens to the atoms in a vapor cell, and the fluorescence of the atoms induced by the pulses was measured with a CCD~(Retiga 3000) through a two-lens relay imaging system with $f_2=50$~mm. We used two interference bandpass filters centered at 780~nm and 794.7~nm, respectively, to measure the fluorescence from the two excited levels, $5P_{1/2}$ and $5P_{3/2}$. The filters had a spectral bandwidth of 3~nm and 50~\% center transmittance.

\section{Results and Discussion}
 \begin{widetext}
  
  \begin{figure}[thb]
   \centerline{\includegraphics[width=0.9\textwidth]{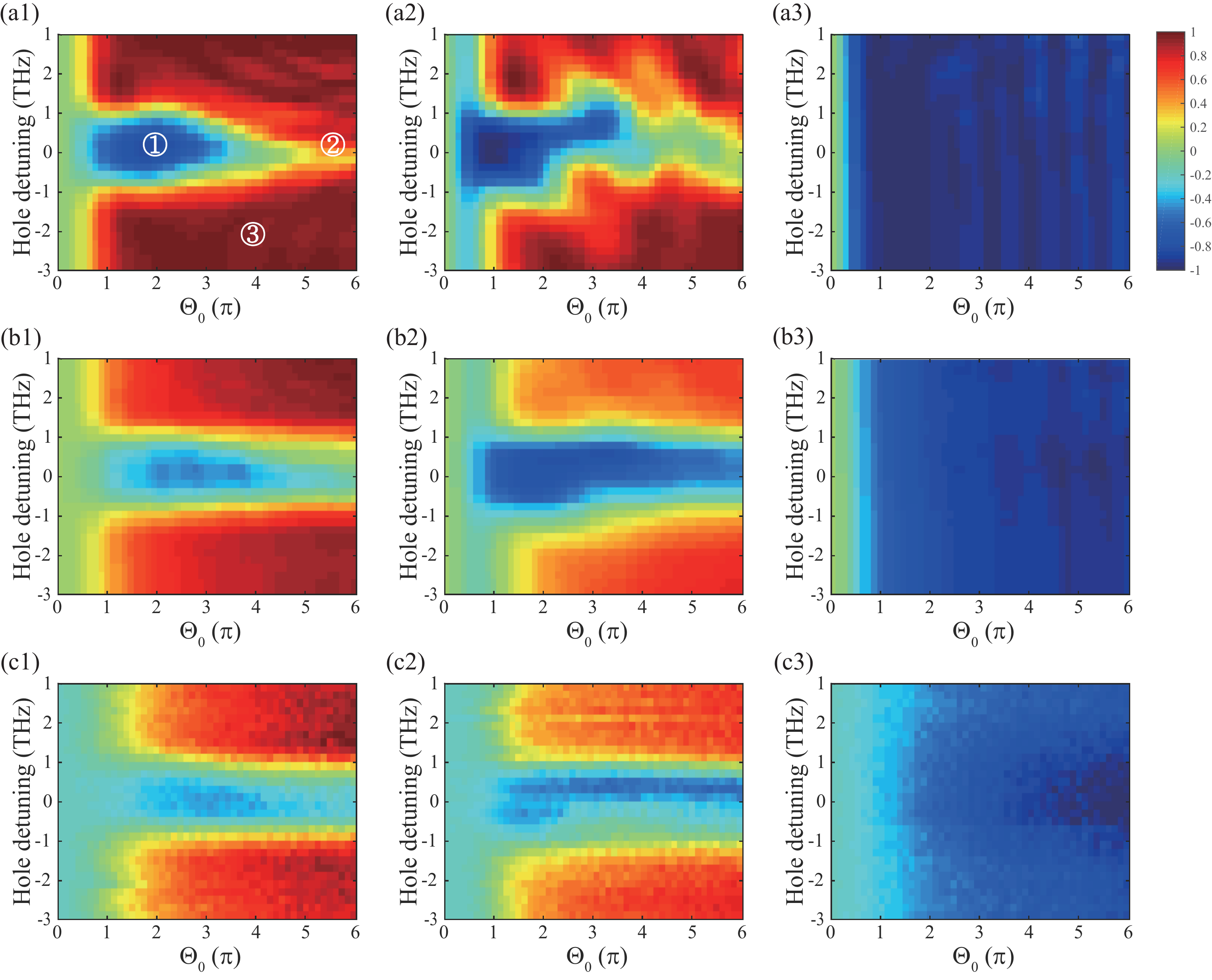}}
    \caption{(Color online) Population difference, $\Delta P = P(\ket{1})-P(\ket{2})$, plotted as a function of the hole detuning, $\omega_h-\omega_1$, and the calibrated pulse area $\Theta_0$: (a) Calculation without the spatial average consideration, (b) Calculation with the spatial average consideration, and (c) Experimental results. The left column (a1, b1, c1)  corresponds to $c_2=50,000~$fs$^2$ (the minimal hole-pulse width condition), the middle column (a2, b2, c2) $c_2=20,000~$fs$^2$ (a long hole-pulse), and the right column (a3, b3, c3) $c_2=-20,000~$fs$^2$ (a negatively-chirped pulse). The color scheme of the figures indicates the final ($t=\infty$) state of the system to be in either $\ket{2}$ (blue) or $\ket{1}$ (red). The calculation used the measured actual laser spectrum, but what used a Gaussian spectrum makes little difference. } 
    \label{fig4}
\end{figure}
 \end{widetext}
 
Figure~\ref{fig4} compares the numerical calculation (a,b) with the experimental results (c). The population difference between the excited states, $\Delta P = P(\ket{1})-P(\ket{2})$,  is plotted for three chirp parameters: $c_2=50,000$~fs$^2$ (the minimal hole-pulse condition), $20,000$~fs$^2$ (a long hole-pulse condition), and $-20,000$~fs$^2$ (a negative chirp). The numerical calculation in Figs.~\ref{fig4}(a1), \ref{fig4}(a2), and \ref{fig4}(a3) shows the chirp-dependent behavior of the given hybrid adiabatic-nonadiabatic interaction. Under the minimal hole-pulse condition in Fig.~\ref{fig4}(a1), a near-zero detuning ($\omega_h \approx \omega_1$) exhibits the as-expected intensity dependence of the selective excitation:  As the pulse area ($\Theta_0$) increases, the state evolves to $\ket{2}$ or $\ket{1}$ in the region marked by \raisebox{.5pt}{\textcircled{\raisebox{-.9pt} {1}}} or \raisebox{.5pt}{\textcircled{\raisebox{-.9pt} {2}}}, respectively. Near \raisebox{.5pt}{\textcircled{\raisebox{-.9pt} {1}}}, the system evolves to $\ket{2}$ through the hybrid adiabatic-nonadiabatic interaction. However, when the adiabatic condition is fully satisfied around \raisebox{.5pt}{\textcircled{\raisebox{-.9pt} {2}}}, the system evolves to $\ket{1}$. Note that the region near \raisebox{.5pt}{\textcircled{\raisebox{-.9pt} {3}}} is the case for a large hole-detuning; this region exhibits an ordinary chirped-RAP behavior, because the far-off-resonant hole plays little role in the overall dynamics. The long hole-pulse case, in Fig.~\ref{fig4}(a2), shows an extended non-adiabatic coupling near both the first and second adiabatic crossing points; thus, the overall dynamics appears to be sensitively dependent on both the hole detuning and the pulse area, as expected. In the negative chirp case, in Fig.~\ref{fig4}(a3), the hole pulse is co-located with the second adiabatic crossing point, resulting in, again, an ordinary chirped-RAP (to $\ket{2}$ in this case because of the negative sign of the chirp), irrespective of the hole detuning.

The second row of Fig.~\ref{fig4} is the spatially-averaged calculation of the first row. Because the laser beam has a Gaussian spatial profile in the transverse direction, each atom in the atom ensemble interacts with a different laser intensity~\cite{LeeOL2015}. The Gaussian beam wait was 250~$\mu$m and the Rayleigh range (about 20~cm) greatly exceeds the size (about 50~$\mu$m) of the imaged area. When this spatial average effect due to the transverse laser beam profile is taken into account, the numerical calculation results in Figs.~\ref{fig4}(b1), \ref{fig4}(b2), and \ref{fig4}(b3), showing good agreement with the experimental result in the third row, Figs.~\ref{fig4}(c1), \ref{fig4}(c2), and \ref{fig4}(c3), respectively.

\begin{figure}[thb]
   \centerline{\includegraphics[width=0.45\textwidth]{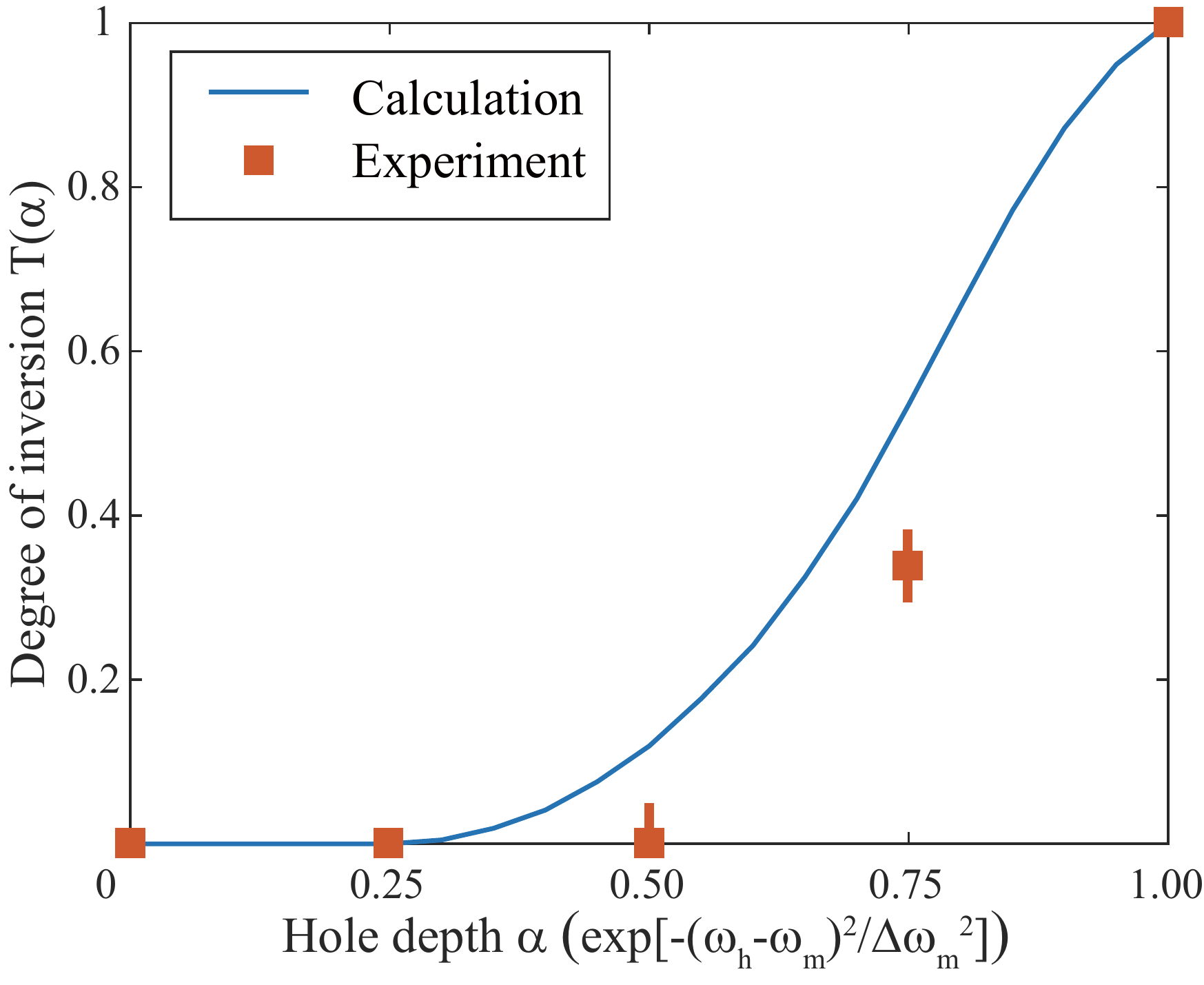}}
    \caption{(Color online) The degree of inversion $T(\alpha)= |\Delta P_{\rm min} (\alpha)/ \Delta P_{\rm min}(\alpha=1.0)|$ as a function of the hole depth ($\alpha$).} 
    \label{fig5}
\end{figure}

Finally, the contribution of the hole to the given selective excitation scheme is shown in Fig.~\ref{fig5}. By changing the hole depth, defined by $\alpha$ in Eq.~\eqref{EComega}, we plot the degree of inversion defined by $T(\alpha)= |\Delta P_{\rm min} (\alpha)/ \Delta P_{\rm min}(\alpha=1.0)|$, where $\Delta P_{\rm min}$ is the minimal population difference $\Delta P$ for a given hole depth $\alpha$ during the entire evolution. When the depth is large, $\alpha \approx \exp[-(\omega_h-\omega_m)^2/\Delta\omega_m^2]$, enough to completely remove the spectrum at $\omega_h$, the ground state evolves to the second excited state (i.e., $\ket{0} \leftrightarrow \ket{2}$) and $\ket{1}$ is unchanged. On the other hand, no hole ($\alpha=0$) induces the ordinary chirped RAP, a cyclic permutation of the energy states. 

We now turn our attention to the implications of the results obtained in this study to possible applications. The first example is the closed adiabatic two-excited state transitions $\ket{1} \leftrightarrow \ket{2}$, which can be made by combining the given hybrid interaction and an ordinary chirped RAP (with a negative chirp): Since an ordinary negatively-chirped RAP induces the cyclic state permutations, a subsequent hybrid interaction $\ket{0} \leftrightarrow \ket{2}$ completes an adiabatic exchange of the excited states, $\ket{1} \leftrightarrow \ket{2}$, and $\ket{0}$ is unchanged. Therefore, with this procedure, an ultrafast time-scale adiabatic control among the excited states of a three-level system may be achieved. The second example is an optical cont
rol of $N$ qubits arranged in a lattice~\cite{BeugnonNP2007,BlattNature2008}. In particular, when a short lattice constant makes an conventional optics with focused beams difficult to address individual qubits, our method may provide a solution: Our calculation (not shown) predicts that spatial beam-shape profiling in conjunction with the given intensity-dependent selective excitation achieves sub-wavelength-scale qubit controls. For example, the atomic qubit gates constructed based on the Rydberg-atom dipole blockade effect often use about 10~$\mu$m-scale optical micro-traps~\cite{9,8,Kim}, so reducing the lattice constant down below one wavelength allows to use significantly lower Rydberg energy states, which may be useful for faster quantum gate operations.
  
\section{Conclusion}
In summary, we have studied the hybrid adiabatic-nonadiabatic quantum dynamics of a three-level system in the $V$-type configuration, implemented with a chirped laser pulse with a spectral hole. Each adiabatic crossing point of the conventional three-level chirped RAP has been found to be individually turned on and off with the spectral hole, enabling selective transition to each excited state by controlling the laser intensity. Experiments have been performed with shaped femtosecond laser pulses and the three lowest energy-levels (5S$_{1/2}$, 5P$_{1/2}$, and 5P$_{3/2}$) of atomic rubidium ($^{85}$Rb), and the result agrees well with the theoretically analyzed dynamics of the three-level system. The result indicates that our method, being combined with the ordinary chirped-RAP, implements an adiabatic transitions between the two excited states. Furthermore the selective excitations by laser intensity control may have applications including selective excitations of atoms or ions arranged in space in conjunction with laser beam profile programming.

\begin{acknowledgements}
This research was supported by Samsung Science and Technology Foundation [SSTF-BA1301-12]. 
\end{acknowledgements}
 \FloatBarrier

\end{document}